\documentclass[letterpaper, 10 pt, journal, twoside]{IEEEtran}

\IEEEoverridecommandlockouts                              

\usepackage{amsmath}																									%
\usepackage{amsfonts}																									%
\usepackage{amssymb}																									%
\usepackage{tcolorbox} 																								
\usepackage{nicefrac}   																									
\usepackage[]{units}																										
\usepackage[latin1]{inputenc}																							%
\usepackage{type1cm}																									%
\usepackage{type1ec}																									%
\usepackage[T1]{fontenc}																								%
\usepackage{dsfont}																										%
\usepackage{booktabs}																									%
\usepackage{array}																										%
\usepackage{enumerate}																								%
\usepackage{makeidx}																									%
\usepackage{fancyhdr}																									%
\usepackage{bm}																											%
\usepackage{lastpage}																									%
\usepackage{multicol}																									%
\usepackage{amsmath, amssymb, amsfonts}																	%
\usepackage{ifthen}																										%
\usepackage{ifpdf}																											%
\usepackage{framed}																										%
\usepackage[amsmath,thmmarks,framed,hyperref]{ntheorem}											%
\DeclareMathAlphabet{\mathpzc}{OT1}{pzc}{m}{it}															%
\usepackage{amsmath,amsfonts,amssymb,booktabs}														%
\usepackage{amstext}																									%
\usepackage{psfrag} 																										%
\usepackage{float} 																										%
\usepackage{cite} 																											%
\usepackage{todonotes}

\usepackage{graphicx} 																									%
\usepackage{textcomp} 																									%
\usepackage{color} 																											%
\usepackage{dsfont} 																										%
\usepackage{tikz}																											%
\usetikzlibrary{matrix}																										%
\usetikzlibrary{arrows,automata,matrix,fit,positioning,calc}             								%
\usepackage{pgfplots}																									%
\usepackage[latin1]{inputenc}																						%
\usepackage{cite}																											%
\usepackage{pgfplots}
\usetikzlibrary{arrows,automata,matrix,fit,positioning,calc}
\usepackage{verbatim}

\usepackage{mathtools}
\usepackage{pst-node}

\usepackage{algorithm}
\usepackage[noend]{algpseudocode}
\makeatletter
\def\BState{\State\hskip-\ALG@thistlm}

\newcommand{\x}{\boldsymbol{x}}
\newcommand{\bu}{\boldsymbol{u}}
\newcommand{\w}{\boldsymbol{w}}
\newcommand{\y}{\boldsymbol{y}}
\newcommand{\bv}{\boldsymbol{v}}
\newcommand{\e}{\boldsymbol{e}}

\newcounter{thmCounter}
\newtheorem{theorem}[thmCounter]{\bfseries Theorem}

\newtheorem{claim}{Claim}

\usepackage{xcolor}

\usepackage{graphicx}

\title{\textsc{DeepCAS}: A Deep Reinforcement Learning Algorithm for Control-Aware Scheduling} 

\author{Burak Demirel, Arunselvan Ramaswamy, Daniel E. Quevedo and Holger Karl
\thanks{A. Ramaswamy was supported by the German Research Foundation (DFG) (project number 315248657).}
\thanks{B. Demirel is with Autonomous Motion, ATS Pre-Development and Research, Scania CV AB (e-mail: burak.demirel@scania.com).}
\thanks{A. Ramaswamy, D. E. Quevedo, and H. Karl are with Paderborn University, Warburger Stra{\ss}e 100, 33098, Paderborn, Germany (e-mail: arunr@mail.uni-paderborn.de, dquevedo@ieee.org, holger.karl@upb.de).} 
}

\begin{document}

\maketitle
\thispagestyle{empty}
\pagestyle{empty} 

\begin{abstract}
We consider networked control systems consisting of multiple independent controlled subsystems, operating over a shared communication network. Such systems are ubiquitous in cyber-physical systems, Internet of Things, and large-scale industrial systems. In many large-scale settings, the size of the communication network is smaller than the size of the system. In consequence, scheduling issues arise. The main contribution of this paper is to develop a deep reinforcement learning-based \emph{control-aware} scheduling (\textsc{DeepCAS}) algorithm to tackle these issues. We use the following (optimal) design strategy: First, we synthesize an optimal controller for each subsystem; next, we design a learning algorithm that adapts to the chosen subsystems (plants) and controllers. As a consequence of this adaptation, our algorithm finds a schedule that minimizes the \emph{control loss}. We present empirical results to show that \textsc{DeepCAS} finds schedules with better performance than periodic ones.
\end{abstract}


\section{Introduction}\label{sec:intro}
\IEEEPARstart{A}{rtificial} intelligence (AI) offers an attractive set of tools that are mostly model-free, yet useful in solving stochastic and optimal control problems arising in cyber-physical systems (CPS), Internet of Things (IoT), and large-scale industrial systems. AI-based solutions have seen a major resurgence in recent years, partly owing to recent advances in computational capacities and owing to advances in deep neural networks for function approximation and feature extraction. Oftentimes, the use of reinforcement learning algorithms or AI in conjunction with traditional controllers reduces the complexity of system design while boosting efficiency.

The abovementioned systems are all characterized by large sizes. However, typical resources, such as communication channels, computational resources, network bandwidth etc., do not scale with system size. In other words, resource allocation is an important problem in this setting. In addition, in a distributed control setting that involves feedback, resource allocation is required to be ``control-aware'', i.e., it is needed to aid in optimizing closed-loop control performance. In such feedback driven systems, controllers often rely on information collected from various sensors to make intelligent decisions. Hence, efficient information dispersion is essential for decision making over communication networks to be effective. As noted earlier, this is a hard problem since the number of communication channels available is much smaller than what is ideally required to transfer data from sensors to controllers. 

\begin{figure}[!t] 
	\centering
	\includegraphics[scale=0.65]{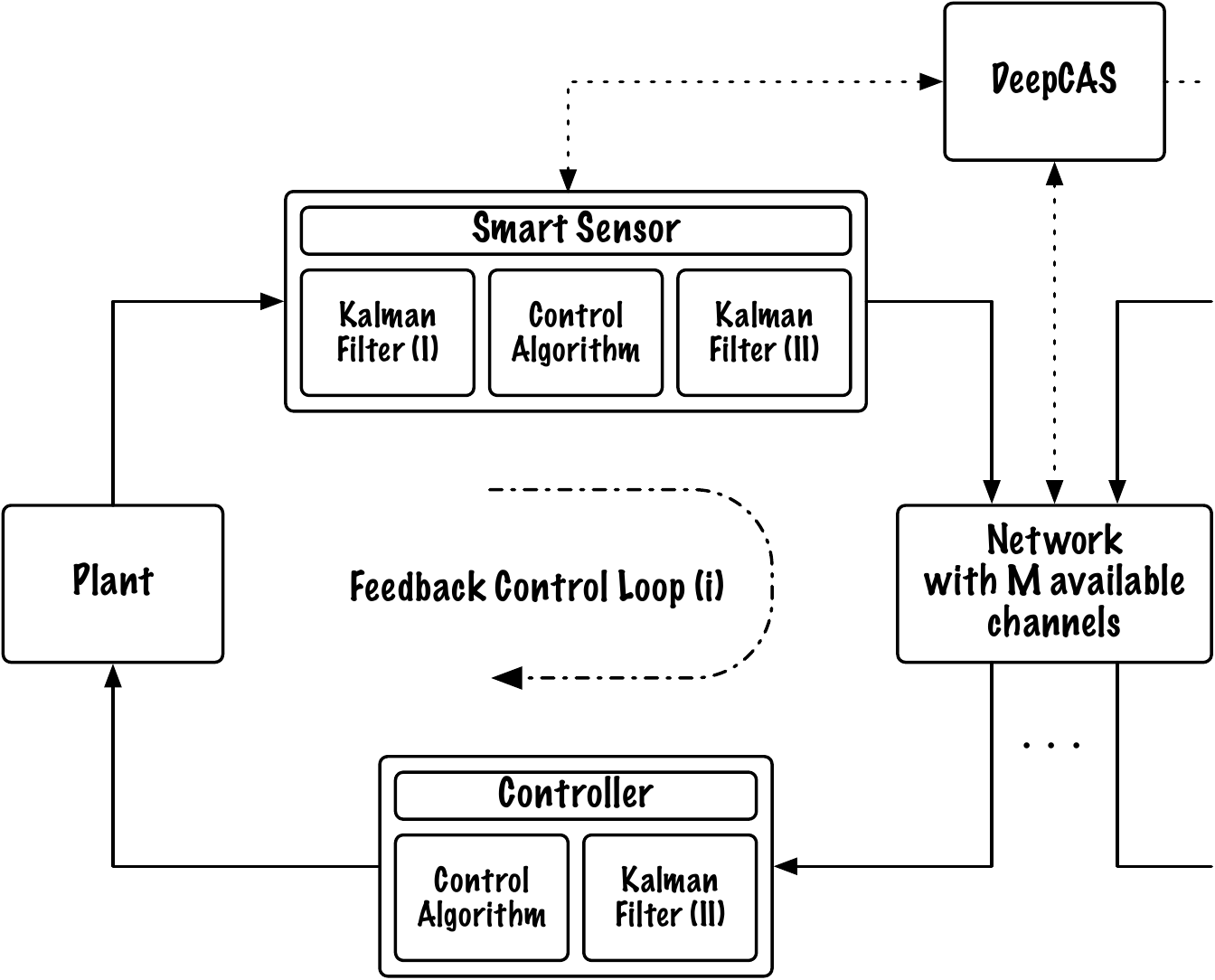}
	\caption{Networked control system (NCS) that consists of $N$ control subsystems closed over a shared communication network.}
	\label{fig:sys_arch}
\end{figure}

Fig.~\ref{fig:sys_arch} illustrates a simplified representation of the class of CPS and IoT systems of interest. The system consists of $N$ \emph{independent} subsystems that communicate over a shared communication network, which contains $M$ channels. We assume that $M \ll N$ ($M$ is much smaller than $N$), and that transmissions are via error-free channels. Each subsystem consists of one smart sensor, one controller, and one plant. Within each subsystem, there is feedback from the sensor to the controller. These feedback loops are closed over this resource-constrained communication network. 

At every stage, \textsc{DeepCAS}, our deep reinforcement learning-based model-free scheduling algorithm, decides which $M$ of the $N$ subsystems are allocated channels to close the feedback loop. \textsc{DeepCAS} takes scheduling decisions by adapting to the control actions while trying to minimize the control loss. At every stage, the smart sensors compute estimates of the subsystem states, using \emph{Kalman Filter (I)}, for transmission to the corresponding controller, see Fig.~\ref{fig:sys_arch}. The controller runs \emph{Kalman Filter (II)} to estimate the subsystem state in the absence of transmissions. In addition to \emph{Kalman Filter (I)}, each smart sensor also implements a copy of \textit{Kalman Filter (II)} and the control algorithm. In other words, the smart sensor is cognizant of the state estimate used by the controller at every time instant. \textsc{DeepCAS} obtains feedbacks (i.e., rewards) from sensors for taking scheduling decisions.

Previously, several scheduling strategies have been proposed to determine the access order of different sensors and/or actuators; see~\cite{PEF+18} and references therein. A popular approach is to use periodic schedules~\cite{ReS:04,HrZ:08,SCC:11,OBG:14} since they are easy to implement and they facilitate stability analysis of networked control systems. Unfortunately, finding optimal periodic schedules for control applications may not be easy since both period and sequence need to be found. Further, restricting to periodic schedules may lead to performance loss~\cite{ZCW+:18}. With a handful of exceptions, the determination of optimal schedules indeed requires solving a mixed-integer quadratic program, which is computationally infeasible for all but very small systems; see~\cite{COZ+:17,ZCW+:18}. 

Event- and self-triggering algorithms present popular alternatives to periodic scheduling; see~\cite{HJT:12} and references therein. Linear, quadratic optimal control problems subject to such scheduling schemes have been investigated in~\cite{RSJ:13,MoH:14,HQP+:15,DLG+:17}. Many of the aforementioned results only consider single-loop control systems. There exists limited literature that study multi-loop control systems~\cite{RSJ:13,MoH:14,HQP+:15}. One limitation is that many of these results only investigate linear scalar systems.

\emph{Our contribution} in the present work is in the development of a deep reinforcement learning-based control-aware scheduling algorithm, \textsc{DeepCAS}. At its heart lies the Deep Q-Network (DQN), a modern variant of Q learning, introduced in~\cite{MKS+:13}. In addition to being readily scalable, \textsc{DeepCAS} is completely model-free. To optimize the overall control performance, we propose the following sequential design of control and scheduling: In the first step, we design an optimal controller for each independent subsystem. As discussed in~\cite{DLG+:17}, under limited communication, the \emph{control loss} has two components: \textbf{(a)} best possible control loss \textbf{(b)} error due to intermittent transmissions. If $M = N$, then \textbf{(b)} vanishes. Since we are in the setting of $M \ll N$, the goal of the scheduler is to minimize \textbf{(b)}. To this end, we first construct an associated Markov decision process (MDP). The state space of this MDP is the difference in state estimates of all controllers and sensors (obtainable from the smart sensors). The single-stage reward is the negative of the loss component \textbf{(b)}. Since we are using DQN to solve this MDP, we do not need the knowledge of transition probabilities. The goal of \textsc{DeepCAS} is to find a scheduling strategy that maximizes the reward, i.e., minimizes \textbf{(b)}.

\section{Networked Control System: Model, Assumptions, and Goals} \label{sec:model_assumptions}
\subsection{Model for each subsystem}
As illustrated in Fig.~\ref{fig:sys_arch}, our networked control system consists of $N$ independent closed-loop subsystems. The feedback loop within each subsystem (plant) is closed over a shared communication network. For $1 \le i \le N$, subsystem $i$ is described by
\begin{align} \label{LTI_eq}
	\x_{t+1}^{(i)} = A_{}^{(i)}\x_{t}^{(i)} + B_{}^{(i)}\bu_{t}^{(i)} + \w_{t}^{(i)} \;,
\end{align}
where $A_{}^{(i)}$ and $B_{}^{(i)}$ are matrices of appropriate dimensions, $\x_{t}^{(i)}\in\mathbb{R}_{}^{n_i}$ is the state of subsystem $i$, $\bu_{t}^{(i)}\in\mathbb{R}_{}^{m_i}$ is the control input, and $\w_{t}^{(i)}\in\mathbb{R}_{}^{n_i}$ is zero-mean i.i.d. Gaussian noise with covariance matrix $W_{}^{(i)}$. The initial state of subsystem $i$, $\x_{0}^{(i)}$, is assumed to be a Gaussian random vector with mean $\bar{\x}_{0}^{(i)}$ and covariance matrix $X_{0}^{(i)}$ and of each other. 

At a given time $t$, we assume that only noisy output measurements are available. We, thus, have:
\begin{align}
	\y_{t}^{(i)} = C_{}^{(i)}\x_{t}^{(i)} + \bv_{t}^{(i)} \;,
\end{align}
where $\bv_{t}^{(i)}\in\mathbb{R}_{}^{p_i}$ is zero-mean i.i.d. Gaussian noise with covariance matrix $V_{}^{(i)}$. The noise sequences, $\w_{t}^{(i)}$ and $v_{t}^{(i)}$, are independent of the initial conditions $\x_{0}^{(i)}$.

\subsection{Control architecture and loss function}
The dynamics of each subsystem is a stochastic linear time-invariant (LTI) system given by~\eqref{LTI_eq}. Further, each subsystem is independently controlled. Dependencies do arise from sharing a communication network. Subsystem $i$ has a smart sensor which samples the subsystem's output $\y^{(i)}_t$ and computes an estimate of the subsystem's state. This value is then sent to the associated controller, provided a channel is allocated to it by \textsc{DeepCAS}. If the controller obtains a new state estimate from the sensor, then it calculates a control command based on this state estimate. Otherwise, it calculates a control command based on its own estimate of the subsystem's state.

The control actions and scheduling decisions (of \textsc{DeepCAS}) are taken to minimize the total control loss given by
\begin{align}
	J = \sum_{i = 1}^{N} J_{}^{(i)}, \label{eq_control_loss}
\end{align}
where $J_{}^{(i)}$ is the expected control loss of subsystem $i$ and is given by
\begin{multline*}
	J_{}^{(i)} = \mathbb{E}\bigg[ \x_{\mathrm{T}}^{(i)\top}Q_{f}^{(i)}\x_{\mathrm{T}}^{(i)} \\ 
	+ \sum_{t=0}^{\mathrm{T}-1}\Big( \x_{t}^{(i)\top}Q_{}^{(i)}\x_{t}^{(i)} + \bu_{t}^{(i)\top}R_{}^{(i)}\bu_{t}^{(i)} \Big) \bigg] \;,
\end{multline*}
where $Q_{}^{(i)}$ and $Q_{f}^{(i)}$ are positive semi-definite matrices and $R_{}^{(i)}$ is positive definite.

\subsection{Smart sensors and pre-processing units}
Within our setting, the primary role of a smart sensor is to take measurements of a subsystem's output. Also, it plays a vital role in helping \textsc{DeepCAS} with scheduling decisions. It is from the smart sensors that \textsc{DeepCAS} gets all the necessary feedback information for scheduling. For these tasks, each smart sensor employs two Kalman filters: (1) \emph{Kalman Filter (I)} is used to estimate the subsystem's state, (2) a copy of \emph{Kalman Filter (II)} is used to estimate the subsystem's state as perceived by the controller. Note that the controller employs \emph{Kalman Filter (II)}. Below, we discuss the set-up in more detail.

\noindent\textbf{Kalman filter (I):} Since we assume that the sensors have knowledge of previous plant inputs, the sensors employ standard Kalman filters to compute the state estimate $\hat{\x}_{t \mid t}^{(i)s}$ and covariance $P_{t\mid t}^{(i)s}$ recursively as:
\begin{align*}
	\hat{\x}_{t \mid t-1}^{(i)s} &= A_{}^{(i)}\hat{\x}_{t-1 \mid t-1}^{(i)s} + B_{}^{(i)}\bu_{t-1}^{(i)} \\
	P_{t \mid t-1}^{(i)s} &= A_{}^{(i)}P_{t-1 \mid t-1}^{(i)s}A_{}^{(i)\top} + W_{}^{(i)} \\
	K_{t}^{} &=  P_{t\mid t-1}^{(i)s}C_{}^{(i)\top}\big( C_{}^{(i)}P_{t\mid t-1}^{(i)s}C_{}^{(i)\top} + V_{}^{(i)} \big)_{}^{-1} \\
	\hat{\x}_{t \mid t}^{(i)s} &= \hat{\x}_{t \mid t-1}^{(i)s} + K_{t}^{(i)}\big( \y_{t}^{(i)} - C\hat{\x}_{t \mid t-1}^{(i)s} \big) \\
	P_{t\mid t}^{(i)s} &= \big( I - K_{t}^{(i)}C_{}^{(i)} \big) P_{t\mid t-1}^{(i)s} \;,
\end{align*}
starting from $\hat{\x}_{0 \mid -1}^{(i)s} = \bar{\x}_{0}^{(i)}$ and $P_{0 \mid -1}^{(i)s} = X_{0}^{(i)}$.

\noindent\textbf{Kalman filter (II):} The controller runs a minimum mean square error (MMSE) estimator to compute estimates of the subsystem's state as follows:
\begin{align}
	\hat{\x}_{t \mid t-1}^{(i)c} & =  A_{}^{(i)} \hat{\x}_{t-1 \mid t-1}^{(i)c} + B_{}^{(i)} \bu_{t-1}^{(i)} \;, \label{eqn:prediction_step_controller} \\
	\hat{\x}_{t \mid t}^{(i)c} & =
	\begin{cases}
		\hat{\x}_{t \mid t}^{(i)s} & \text{if the MMSE estimate received}\;, \\
		\hat{\x}_{t \mid t-1}^{(i)c} & \text{otherwise} \;,
	\end{cases} 
\end{align}
with $\hat{\x}_{0 \mid -1}^{(i)c} = \bar{\x}_{0}^{(i)}$.

\subsection{Goal: minimizing the control loss} \label{sec_control_loss}
For the control problem studied, the certainty equivalent (CE) controller is still optimal; see~\cite{DLG+:17} for details. Using the control commands, generated by the CE controllers, the minimum value of the total control loss, \eqref{eq_control_loss}, has two components: \textbf{(a)} best possible control loss \textbf{(b)} error due to intermittent communications. Hence, the problem of minimizing control loss has two separate components: (i) designing the best (optimal) controller for each subsystem and (ii) scheduling in a control-aware manner.  

\noindent\textit{\textbf{Component I: Controller design.}}
The controller in feedback loop $i$ takes the following control action, $\bu_{t}^{(i)}$, at time $t$: 
\begin{align} \label{eq_control_action}
	\bu_{t}^{(i)} = - L_{t}^{(i)} \hat{\x}_{t \mid t}^{(i)c},
\end{align}
where $\hat{\x}_{t \mid t}^{(i)c}$ is the state estimate used by the controller,
\begin{align}
	L_{t}^{(i)} = (B_{}^{(i)\top}S_{t+1}^{(i)}B_{}^{(i)} + R_{}^{(i)})_{}^{-1} B_{}^{(i)\top}S_{t+1}^{(i)}A_{}^{(i)}  \label{eqn:optimal_control_gain}
\end{align}
and $S_{t}^{(i)}$ is recursively computed as
\begin{multline}
	S_{t}^{(i)} = A_{}^{(i)\top} S_{t+1}^{(i)} A_{}^{(i)} + Q_{}^{(i)} - A_{}^{(i)\top}S_{t+1}^{(i)}B_{}^{(i)} \\ \times (B_{}^{(i)\top}S_{t+1}^{(i)}B_{}^{(i)} + R_{}^{(i)})_{}^{-1} B_{}^{(i)\top}S_{t+1}^{(i)}A_{}^{(i)} , \label{eqn:riccati_equation}
\end{multline}
with initial values $S_{N}^{(i)} = Q_{f}^{(i)}$. Let $\hat{\x}_{t \mid t}^{(i)s}$ be the state estimate
of \textit{Kalman Filter (I)}, as employed by the sensor. We have $\hat{\x}_{t \mid t}^{(i)c} =  \hat{\x}_{t \mid t}^{(i)s}$ when the sensor and controller of the feedback loop $i$ have communicated. Otherwise, $\hat{\x}_{t \mid t}^{(i)c}$ is the state estimate obtained from \textit{Kalman Filter (II)}. The minimum value of the control loss of subsystem $i$ is given by
\begin{multline}
		J_{}^{(i)} =\; \bar{\x}_{0}^{(i)\top}S_{0}^{(i)}\bar{\x}_{0}^{(i)} + \textnormal{Tr}\big( S_{0}^{(i)}X_{0}^{(i)} \big) + \sum_{t=0}^{\mathrm{T}-1}\textnormal{Tr}\big( S_{t+1}^{(i)}W_{}^{(i)} \big) \\
		+ \sum_{t=0}^{\mathrm{T}-1} \textnormal{Tr}\big( P_{t \mid t}^{(i)s} \Gamma_{t}^{(i)} \big) + \sum_{t=0}^{\mathrm{T}-1} \mathbb{E}\Big[ \e_{t \mid t}^{(i)\top} \Gamma_{t}^{(i)} \e_{t \mid t}^{(i)} \Big] \;, \label{eqn:inf_expected_cost} 
\end{multline}
where $\Gamma_{t}^{(i)} \triangleq L_{t}^{(i)\top}(B_{}^{(i)\top}S_{t+1}^{(i)}B_{}^{(i)} + R_{}^{(i)})L_{t}^{(i)}$ and $\e_{t \mid t}^{(i)} \triangleq \hat{\x}_{t \mid t}^{(i)s} - \hat{\x}_{t \mid t}^{(i)c}$ stems from communication errors in subsystem $i$. Recall that there are $N$ subsystems and $M << N$ communication channels.

\noindent\textit{\textbf{Component II: Control-aware scheduling.}}
The main aim of the scheduling algorithm is to help minimize $J$ of (\ref{eq_control_loss}). To this end, one must minimize
\begin{align}
	 \sum_{t=0}^{\mathrm{T}-1} \mathbb{E}\Big[ \e_{t \mid t}^{(i)\top} \Gamma_{t}^{(i)} \e_{t \mid t}^{(i)} \Big] \label{eqn:sum_error}
\end{align}
of~\eqref{eqn:inf_expected_cost} for every $1 \le i \le N$. Note that $\mathrm{T}$ in~\eqref{eqn:sum_error} is the control horizon. At any time $t$, the scheduler decides which $M$ among the $N$ subsystems may communicate. Note that $\e_{t \mid t}^{(i)} = 0$ when a communication channel is assigned to subsystem $i$ at time~$t$.

In the following section, we present a deep reinforcement learning algorithm for control-aware scheduling called \textsc{DeepCAS}. \textsc{DeepCAS} communicates only with the smart sensors. At every time instant, sensors are told if they can transmit to their associated controllers. Then, the sensors provide feedback on the scheduling decision for that stage. Note that we do not consider the overhead involved in providing feedback. 

\section{Deep reinforcement learning for control-aware scheduling}\label{sec:mdp}
As stated earlier, at the heart of \textsc{DeepCAS} lies the DQN. The DQN is a modern variant of Q-learning that effectively counters Bellman's curse of dimensionality. Essentially, DQN or Q-learning finds a solution to an associated Markov decision process (MDP) in an iterative model-free manner. Before proceeding, let us recall the definition of an MDP. For a more detailed exposition, the reader is referred to~\cite{BeT:96}. An MDP, $\mathcal{M}$, is given by the following tuple $(\mathcal{S}, \mathcal{A}, P, r, \gamma)$, where
\begin{itemize}
 \item[$\mathcal{S}$] is the state-space of $\mathcal{M}$;
 \item[$\mathcal{A}$] is the set of actions that can be taken;
 \item[$P$] is the transition probability, i.e., $P(s, s'; a)$ is the probability of transitioning to state $s'$ when action $a$ is taken at state $s$;
 \item[$r$] is the one stage reward function, i.e., $r(s, a)$ is the reward when action $a$ is taken at state $s$;
 \item[$\gamma$] is the discount factor with $\gamma \in [0,1]$.
\end{itemize}
Below, we state the MDP $\mathcal{M}_{d}^{}$ associated with our problem.
\begin{itemize}
 \item[$\mathcal{S}$:] The state space $\mathcal{S}$ consists of all possible \textit{augmented error vectors}. Hence, the state vector $s_t$ at time $t$ is given by $(\e_{t \mid t}^{(1)}, \ldots, \e_{t \mid t}^{(N)})$.
 \item[$\mathcal{A}$:] Action space is given by the $M$-size subsets of the channels: $\left\{\mathcal{S} \mid \mathcal{S} \subset \{1, 2, \ldots, N\},\ \lvert \mathcal{S} \rvert = M \right\}$. Hence, the cardinality of the action space is given by $|\mathcal{A}| = \binom{N}{M}$.
 \item[$r$:] At time $t$, the reward $r(t)$ is given by $- \sum_{i=1}^{N} \e_{t \mid t}^{(i)} \Gamma_{t}^{(i)} \e_{t \mid t}^{(i)} $.
 \item[$\gamma$:] Although it would seem natural to use $\gamma = 1$, we use $0 < \gamma < 1$ since it hastens the rate of convergence.
\end{itemize}
Note that the scheduler (\textsc{DeepCAS}) takes action just before time $t$ and receives rewards just after time $t$, based on transmissions at time $t$. Also, note that \textsc{DeepCAS} only gets non-zero rewards from non-transmitting sensors. \textsc{DeepCAS} is model-free. Hence, it does not need to know transition probabilities. 

Let us suppose we use a reinforcement learning algorithm, such as Q-learning, to solve $\mathcal{M}_d$. Since the learning algorithm will find policies that minimize the future expected cumulative rewards, we expect to find policies that minimize scheduling effects on the entire system. This is a consequence of our above definition of reward $r$. Below, we provide a brief overview of Q-learning and DQN, the reinforcement learning algorithm at the heart of \textsc{DeepCAS}. \emph{Simply put, \textsc{DeepCAS} is a DQN solving the above defined MDP $\mathcal{M}_d$.}

\noindent\textbf{\textsc{DeepCAS}.}
At any time $t_0$, the scheduler is interested in maximizing the following expected discounted future reward:
\[
 R(t_0) := \mathbb{E} \left[ \sum \limits_{t = t_0}^{T-1} \gamma^{t - t_0} r(t) \right].
\]
Recall that $r(t)$ is the single stage cost given by $- \sum \limits_{i=1}^{N} \e_{t \mid t}^{(i)} \Gamma_{t}^{(i)} \e_{t \mid t}^{(i)} $. $Q$-learning is a useful methodology to solve such problems. It is based on finding the following Q-factor for every state-action pair:
\begin{align*}
	 Q^*(s,a) := \max _\pi \mathbb{E} \left[R_t \mid s_t = s, a_t = a, \pi \right],
\end{align*}
where $\pi$ is a policy that maps states to actions. The algorithm itself is based on the Bellman equation:
\begin{align*}
	Q^*(s, a) = \mathbb{E}_{s' \sim \mathcal{E}} [r + \gamma \max \limits_{a' \in \mathcal{A}} Q^*(s', a') \mid s,a] \;.
\end{align*}
Note that \textsc{DeepCAS} has no knowledge of networked control system dynamics. This unknown dynamics is represented by $\mathcal{E}$, in the above equation.
Since our state space is continuous, we use a deep neural network (DNN) for function approximation. Specifically, we try to find good approximations of the Q-factors iteratively. In other words, the neural network takes as input state $s$ and outputs $Q(s,a, \theta)$ for every possible action $a$, such that $Q(s,a, \theta) \approx Q^*(s,a)$. This deep function approximator, with weights $\theta$, is referred to as a Deep Q-Network. The Deep Q-Network is trained by minimizing a time-varying sequence of loss functions $L_t (\theta_t)$ given by
\begin{align*}
	 L_t (\theta_t) = \big(\nicefrac{1}{2}\big)\;\mathbb{E}_{s,a \sim \rho(s,a)} \left[ (Q(s,a, \theta_t) - y_t)^2 \right],
\end{align*}
where $y_t := \mathbb{E}_{s' \sim \mathcal{E}} \left[ r + \gamma \max_{a'} Q(s', a', \theta_{t-1}) \mid s,a \right]$ is the expected cost-to-go based on the latest update of the weights; $\rho$ is the behavior distribution \cite{MKS+:13}. Training the neural network involves finding $\theta^*$, which minimizes the loss functions. Since the algorithm is run online, training is done in conjunction with scheduling. At time $t$, after feedback (reward) is received, one gradient descent step can be performed using the following gradient term:
\begin{multline}
 \nabla_{\theta_t} L_t (\theta_t) = \mathbb{E}_{s,a \sim \rho(\cdotp); s \sim \mathcal{E}}
\bigg[ \Big( Q(s,a; \theta_t) - r \\
 - \gamma \max_{a'} Q(s', a', \theta_{t-1}) \Big) \nabla_{\theta_t} Q(s,a; \theta_t) \bigg]. \label{eq_grad_desc}
\end{multline}
To make the algorithm implementable, we update the weights $\theta_t$ using samples than finding the above expectation exactly. At each time, we pick actions using the $\epsilon$-greedy approach \cite{MKS+:13}. Specifically, we pick a random action with probability $\epsilon$, and we pick a greedy action with probability $1 - \epsilon$. This $\epsilon$-greedy approach for picking actions induces the behavior distribution $\rho$. In other words, the actions at every stage are picked using distribution $\rho$. Note that a greedy action $a_t$ at time $t$ is one that  maximizes $Q(s_t, a; \theta)$. Initially it is desirable to \textit{explore}, hence $\epsilon$ is set to $1$. Once the algorithm has gained some experience, it is better to \textit{exploit} this experience. To accomplish this, we use an attenuating $\epsilon$ to $0$.

Although we train our DNN in an online manner, we do not perform a gradient descent step using~\eqref{eq_grad_desc}, since it can lead to poor learning. Instead, we store the previous $K$ experiences $(s_t, a_t, r_t, s_{t+1})$,  $t_0 - K + 1 \le t \le t_0 $, in an \textit{experience replay memory} $\mathcal{D}$. When it comes to training the neural network at time $t$,  it performs a single mini-batch gradient descent step. The mini-batch (of gradients) is randomly sampled from the aforementioned experience replay $\mathcal{D}$. The idea of using experience replay memory, to overcome biases and to have a stabilizing effect on algorithms, was introduced in~\cite{MKS+:13}.

\begin{algorithm}{DQN for control-aware scheduling}
\begin{algorithmic}[1]
\State Initialize the replay memory $\mathcal{D}$ to capacity $K$.
\State Initialize the weights, $\theta$, of the Q-Network. 
\For{ the entire duration}
\State With probability $\epsilon$ select a random action $a_t$.
\State With probability $1 - \epsilon$ pick $a_t$ that maximizes $Q(s_t, a, \theta)$.
\State Execute action $a_t$ to obtain reward $r_t$ and observe $s_{t+1}$.
\State Store $(s_t, a_t, r_t, s_{t+1})$ in $\mathcal{D}$.
\State Sample random mini-batch transitions ($(s_j, a_j, r_j, s_{j+1})$) from $\mathcal{D}$.
\State Corresponding to $(s_j, a_j, r_j, s_{j+1})$, set 
\[ 
y_j := r_j + \gamma \max_{a'} Q(s_{j+1},a'; \theta).
 \]
\State Perform a gradient descent step with loss given by $(y_j - Q(s_j, a_j; \theta))^2$.
\EndFor
\end{algorithmic}
\end{algorithm}

\section{Experimental results} \label{sec:numerical_example}
Recall that DQN is at the heart of our \textsc{DeepCAS}, which uses a deep neural network to approximate Q-factors. The input to this neural network is the appended error vector. The hidden layer consists of 1024 rectifier units. The output layer is a fully connected linear layer with a single output for each of the ${N \choose M}$ actions. The discount factor $\gamma$ in our Q-learning algorithm is fixed at $0.95$. The size of the experience replay buffer is fixed at $20,000$. The exploration parameter $\epsilon$ is initialized to $1$, then attenuated to $0.001$ at the rate of $0.9$. For training the neural network, we use the optimizer ADAM~\cite{KiB:15} with a learning rate of $\e^{-4}$ and a decay of $0.001$. The control horizon is set to $T = 500$. Note that we used the same set of parameters for all of the experiments presented below.

\textit{We conducted three sets of experiments. For the first two sets, we used the reward described in Section~\ref{sec:mdp}. For the last experiment, we used the total control cost as the reward. The reader is referred to \eqref{eqn:inf_expected_cost} in Section~\ref{sec_control_loss} for the control cost associated with subsystem $i$. Using the full control cost as the reward allows us to discuss the stability of the networked control system, see Section~\ref{sec:discussion_stability} for details.}

\subsection{Experiment 1 (N=$3$, M=$1$, and T=$500$)}
For our first experiment, we used \textsc{DeepCAS} to schedule one channel for three subsystems. We considered three second-order single-input-single-output (SISO) subsystems consisting of one stable (subsystem $2$) and two unstable subsystems (subsystems $1$ and $3$). If there were three channels, then there would be no scheduling problem and the total optimal control loss $J$ would be $13.8487$. Since there is only a single channel available, one expects a solution to the scheduling problem to allocate it to subsystems $1$ and $3$ for a more substantial fraction of the time, as compared to subsystem $2$. This expectation is fair since subsystems $1$ and $3$ are unstable while subsystem $2$ is stable. \emph{Once trained, on an average \textsc{DeepCAS} indeed allocates the channel to subsystem 1 for 52\% of the time, to subsystem 2 for 12\% of the time, and to subsystem 3 for 36\% of the time.}

\begin{figure}[!h] 
	\centering
	\includegraphics[scale=1.0]{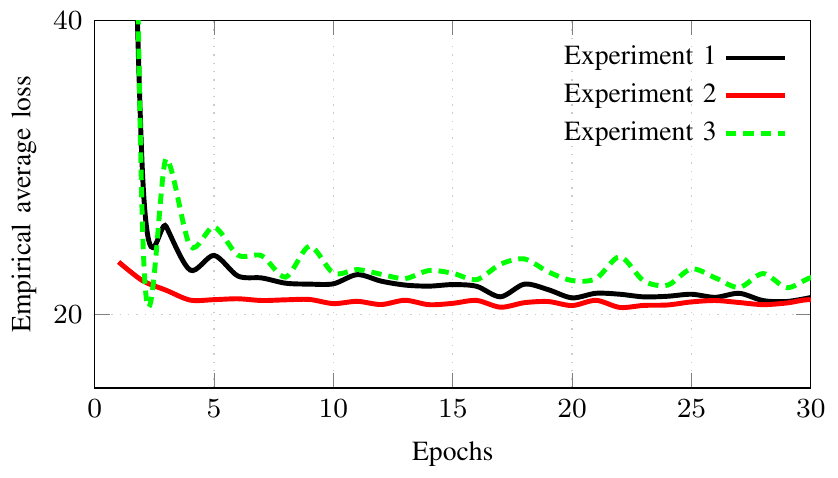}
  \caption{Convergence of the empirical average control loss.}
	\label{fig:three_plants_error}
\end{figure}

We train \textsc{DeepCAS} continuously over many epochs. Each epoch corresponds to a single run of the control problem with horizon $500$. At the start of each epoch, the initial conditions for the control problem are chosen as explained in \S~\ref{sec:model_assumptions}. The \textit{black-curve} in Fig.~\ref{fig:three_plants_error} illustrates the learning progress in \emph{Experiment 1}. The abscissa axis of the graph represents the epoch number while the ordinate axis represents the average control loss. The plot is obtained by taking the mean of $30$ Monte Carlo runs. Since DQN is randomly initialized, scheduling decisions are poor at the beginning, and the average control loss is high. As learning proceeds, the decisions taken improve. After only $10$ epochs, \textsc{DeepCAS} converges to a scheduling strategy with an associated control loss of around~$21$.

Traditionally, the problem of scheduling for control systems is solved by using control theoretic heuristics to find periodic schedules. For \emph{Experiment 1}, we exhaustively searched the space of all periodic schedules, with periods ranging from $2$ to $11$. Using this strategy, we were able to acheive a minimum possible control loss of $J = 22.8112$. In comparison, \textsc{DeepCAS} finds a scheduling strategy with an associated control loss of $21.15$. \emph{In addition to being faster, \textsc{DeepCAS} does not need any system specification and can schedule efficiently for very long control horizons.}

\subsection{Experiment 2 (N=$6$, M=$3$, and T=$500$)}
For our second experiment, we train \textsc{DeepCAS} to schedule three channels for a system with six second-order SISO subsystems. If $N=M$, then the total control loss would be $18.234$. As before, learning is done continuously over many epochs. The \textit{red-curve} in Fig.~\ref{fig:three_plants_error} illustrates the learning progress of \textsc{DeepCAS} in scheduling three channels among six subsystems. The abscissa and ordinate axes are as before. As evidenced in the figure, \textsc{DeepCAS} quickly finds schedules with an associated control loss of around $20$.

We are unable to compare the results of \emph{Experiment 2} with any optimal periodic schedules. This is because optimal periodic scheduling strategies do not extend to the system size and control horizon considered here. Further, performing an exhaustive search for finding periodic schedules is not possible since the number of possibilities are in the order of ${6 \choose 3}^{n} = 20^{n}$, where $n$ is the period-length.

\subsection{Experiment 3 (same set-up as Experiment 1 but with $-J$ as reward)}
The systems considered hitherto have independent subsystems. This facilitates the splitting of the total control cost into two components; see~\eqref{eqn:inf_expected_cost}. The one-stage reward in our algorithm is the negative of the error due to lack of communication defined in~\eqref{eqn:sum_error}. However, in general multi-agent settings, the previously mentioned splitting may not be possible. To show that our results are readily extensible to more general settings, we repeated \emph{Experiments} 1 and 2 with \emph{negative of the one-stage control cost} as the reward. The results of the modified experiments are very similar to the original ones. The learning progress of the modified \emph{Experiment 1}, with full cost, is given by the \textit{green-curve} in Fig.~\ref{fig:three_plants_error}.

\section{Stability issues} \label{sec:discussion_stability}
In our framework, the controller and scheduler run in tandem. The control policy, $\pi_{c}^{}$, is fixed before the scheduler is trained. As a consequence of training, the scheduler finds a scheduling policy $\pi_{s}^{}$. Thus, the controller-scheduler pair finds a policy tuple $(\pi_c, \pi_s)$. To investigate the stabilizing properties of \textsc{DeepCAS}, we make the following mild assumptions on this policy tuple.
\begin{itemize}
	\item[\textbf{A1}] $\liminf \limits_{n \to \infty} \tfrac{1}{n} \sum_{t=0}^n J(t) = \limsup \limits_{n \to \infty} \tfrac{1}{n} \sum_{t=0}^n J(t)$, where $J(t) := \sum_{i=1}^N J_{}^{(i)}(t)$ is the single-stage control loss and 
	\begin{align*}
		J_{}^{(i)}(t) =\;& \bar{\x}_{0}^{(i)\top}S_{0}^{(i)}\bar{\x}_{0}^{(i)} + \textnormal{Tr}\big( S_{0}^{(i)}X_{0}^{(i)} \big) + \textnormal{Tr}\big( S_{t+1}^{(i)}W_{}^{(i)} \big) \\
		&+  \textnormal{Tr}\big( P_{t \mid t}^{(i)s} \Gamma_{t}^{(i)} \big) + \mathbf{E}\Big[ \e_{t \mid t}^{(i)\top} \Gamma_{t}^{(i)} \e_{t \mid t}^{(i)} \Big]
	\end{align*}
	is the single stage loss of subsystem $i$ at time $t$. In other words, we assume that the limit of the average cost sequence exists. This limit may be infinite, i.e., $\lim \limits_{n \to \infty} \tfrac{1}{n} \sum_{t=0}^n J(t) < \infty ~ \text{or} ~ = \infty$.
	\item[\textbf{A2}] The discount factor $\gamma$ used for training is such that $\liminf \limits_{\alpha \uparrow 1} (1 - \alpha) \sum \limits_{t=0}^\infty \alpha ^t J(t) \le \sum \limits_{t=0}^\infty \gamma ^t J(t) + M_0$, for some $0 < M_0 < \infty$. Again, it could be that $\sum_{t=0}^\infty \gamma^t J(t) = \infty$. In which case, \textbf{(A2)} is trivially satisfied.
\end{itemize}

In our framework, the controller uses a control policy, $\pi_c$, that solves the average cost control problem. The scheduler learns a scheduling policy, $\pi_s$, to solve the discounted cost problem. Since they run in tandem, the control loss value $J(t)$, at any time $t$, depends on both the control and scheduling actions taken at time $t$. \textit{Further, we have empirically observed that our scheduler can be successfully trained for all discount factors $\gamma$ close to $1$}. Before proceeding, consider the following theorem due to Abel:
\begin{theorem}[Abel, \cite{LOL:12}]
 Let $\{c_t\}_{t \ge 0}$ be a sequence of positive real numbers, then
 \[
  \liminf \limits_{n \to \infty} \tfrac{1}{n} \sum c_t \le \liminf \limits_{\alpha \uparrow 1} (1 - \alpha) \sum \limits_{t=0}^\infty \alpha ^t c_t.
 \]
\end{theorem}
It follows from \textbf{(A1)} and Abel's theorem that 
\[
\lim \limits_{n \to \infty} \tfrac{1}{n} \sum J(t) \le \liminf \limits_{\alpha \uparrow 1} (1 - \alpha) \sum \limits_{t=0}^\infty \alpha ^t J(t) \;.
\] 
Recall that our scheduler can be successfully trained to solve the discounted cost problem for all discount factors close to (but not equal to) $1$. In other words, given a discount factor $\gamma \approx 1$, the scheduler finds a policy $\pi_s(\alpha)$ such that 
\[
 \sum \limits_{t=0}^\infty \gamma ^t J(t) < \infty.
\]
If we couple this observation with \textbf{(A2)}, we get:
\[
 \liminf \limits_{\alpha \uparrow 1} (1 - \alpha) \sum \limits_{t=0}^\infty \alpha ^t J(t) \le \sum \limits_{t=0}^\infty \gamma ^t J(t) + M(\gamma) < \infty,
\]
for some $\gamma \approx 1$ and $M(\gamma) > 0$. If we choose $\gamma$ as the discount factor for our training algorithm, it follows that:
\begin{align*}
 \lim \limits_{n \to \infty} \tfrac{1}{n} \sum J(t) &\le \liminf \limits_{\alpha \uparrow 1} (1 - \alpha) \sum \limits_{t=0}^\infty \alpha ^t J(t) \nonumber\\
 &\le \sum \limits_{t=0}^\infty \gamma^{t} J(t) + M(\gamma) < \infty. 
\end{align*}
We claim that system stability follows from this set of inequalities. To see this, observe that $\sum_{i=1}^N \lVert \x_{t}^{(i)} \rVert_Q \le J(t)$. Hence, $\limsup \limits_{n \to \infty} \tfrac{1}{n} \sum_{t=0}^n \sum_{i=1}^N\lVert \x_{t}^{(i)} \rVert_Q \le \lim \limits_{n \to \infty} \tfrac{1}{n} \sum_{t=0}^{n} J(t) < \infty$. In other words, the following claim is immediate.

\begin{claim}
	Under \textbf{(A1)} and \textbf{(A2)}, the scheduling algorithm can be successfully trained for discount factors close to $1$, consequently $\sum_{t=0}^\infty \gamma_{}^{t} J(t) < \infty$. Further, the policy $ \pi_{s}$ thus found, stabilizes the system, \textit{i.e.,} $ \sup \limits_{t \ge 0} \sum_{i=1}^{N} \lVert \x_{t}^{(i)} \rVert_Q < \infty$.
\end{claim}

\section{Conclusions} \label{sec:conclusion}
This paper considered the problem of scheduling the sensor-to-controller communication in a networked control system, consisting of multiple independent subsystems. To this end, we presented \textsc{DeepCAS}, a reinforcement learning-based control-aware scheduling algorithm. This algorithm is model-free and scalable, and it outperforms scheduling heuristics, such as periodic schedules, tailored for feedback control applications.

\bibliographystyle{IEEEtran}
\bibliography{bibliography}

\begin{thebibliography}{10}
\providecommand{\url}[1]{#1}
\csname url@samestyle\endcsname
\providecommand{\newblock}{\relax}
\providecommand{\bibinfo}[2]{#2}
\providecommand{\BIBentrySTDinterwordspacing}{\spaceskip=0pt\relax}
\providecommand{\BIBentryALTinterwordstretchfactor}{4}
\providecommand{\BIBentryALTinterwordspacing}{\spaceskip=\fontdimen2\font plus
\BIBentryALTinterwordstretchfactor\fontdimen3\font minus
  \fontdimen4\font\relax}
\providecommand{\BIBforeignlanguage}[2]{{%
\expandafter\ifx\csname l@#1\endcsname\relax
\typeout{** WARNING: IEEEtran.bst: No hyphenation pattern has been}%
\typeout{** loaded for the language `#1'. Using the pattern for}%
\typeout{** the default language instead.}%
\else
\language=\csname l@#1\endcsname
\fi
#2}}
\providecommand{\BIBdecl}{\relax}
\BIBdecl

\bibitem{PEF+18}
P.~Park, S.~C. Ergen, C.~Fischione, C.~Lu, and K.~H. Johansson, ``Wireless
  network design for control systems: a survey,'' \emph{IEEE Communications
  Surveys \& Tutorials}, vol.~20, no.~2, pp. 978 -- 1013, Secondquarter 2018.

\bibitem{ReS:04}
H.~Rehbinder and M.~Sanfridson, ``Scheduling of a limited communication channel
  for optimal control,'' \emph{Automatica}, vol.~40, no.~3, pp. 491--500, March
  2004.

\bibitem{HrZ:08}
D.~Hristu-{V}arsakelis and L.~Zhang, ``{LQG} control of networked control
  systems,'' \emph{International Journal of Control}, vol.~81, no.~8, pp.
  1266--1280, 2008.

\bibitem{SCC:11}
L.~Shi, P.~Cheng, and J.~Chen, ``Optimal periodic sensor scheduling with
  limited resources,'' \emph{IEEE Transactions on Automatic Control}, vol.~56,
  no.~9, pp. 2190--2195, 2011.

\bibitem{OBG:14}
L.~Orihuela, A.~Barreiro, F.~G\'omez-{E}stern, and F.~R. Rubio, ``Periodicity
  of {K}alman-based scheduled filters,'' \emph{IEEE Transactions on Automatic
  Control}, vol.~50, no.~10, pp. 2672--2676, 2014.

\bibitem{ZCW+:18}
M.~Zanon, T.~Charalambous, H.~Wymeersch, and P.~Falcone, ``Optimal scheduling
  of downlink communication for a multi-agent system with a central observation
  post,'' \emph{IEEE Control Systems Letters}, vol.~2, no.~1, pp. 37--42, Jan.
  2018.

\bibitem{COZ+:17}
T.~Charalambous, A.~Ozcelikkale, M.~Zanon, P.~Falcone, and H.~Wymeersch, ``On
  the resource allocation problem in wireless networked control systems,'' in
  \emph{Proceedings of the $57^{th}$ IEEE Conference on on Decision and
  Control}, 2017.

\bibitem{HJT:12}
W.~Heemels, K.~H. Johansson, and P.~Tabuada, ``An introduction to
  event-triggered and self-triggered control,'' in \emph{Proceedings of the
  $51^{st}$ IEEE Conference on Decision and Control}, Dec. 2012.

\bibitem{RSJ:13}
C.~Ramesh, H.~Sandberg, and K.~H. Johansson, ``Design of state-based schedulers
  for a network of control loops,'' \emph{IEEE Transactions on Automatic
  Control}, vol.~58, no.~8, pp. 1962--1975, Aug. 2013.

\bibitem{MoH:14}
A.~Molin and S.~Hirche, ``Price-based adaptive scheduling in multi-loop control
  systems with resource constraints,'' \emph{IEEE Transactions on Automatic
  Control}, vol.~59, no.~12, pp. 3282--3295, Dec. 2014.

\bibitem{HQP+:15}
E.~Henriksson, D.~E. Quevedo, H.~Sandberg, and K.~H. Johansson, ``Multiple loop
  self-triggered model predictive control for network scheduling and control,''
  \emph{IEEE Transactions on Control Systems Technology}, vol.~23, no.~6, pp.
  2167--2181, 2015.

\bibitem{DLG+:17}
B.~Demirel, A.~S. Leong, V.~Gupta, and D.~E. Quevedo, ``Trade-offs in
  stochastic event-triggered control,'' \emph{arXiv:1708.02756}, 2017.

\bibitem{MKS+:13}
V.~Mnih, K.~Kavukcuoglu, D.~Silver, A.~Graves, I.~Antonoglou, D.~Wierstra, and
  M.~Riedmiller, ``Playing atari with deep reinforcement learning,'' in
  \emph{NIPS Deep Learning Workshop}, 2013.

\bibitem{BeT:96}
D.~P. Bertsekas and J.~N. Tsitsiklis, \emph{Neuro-Dynamic Programming}.\hskip
  1em plus 0.5em minus 0.4em\relax Athena Scientific, 1996.

\bibitem{KiB:15}
D.~P. Kingma and J.~Ba, ``Adam: A method for stochastic optimization,'' in
  \emph{Proceeding of the $3^{rd}$ International Conference for Learning
  Representations}, 2015.

\bibitem{LOL:12}
O.~Hernandez-Lerma and J.~B. Lasserre., \emph{Discrete-time Markov control
  processes: basic optimality criteria}.\hskip 1em plus 0.5em minus 0.4em\relax
  Springer Science \& Business Media, 2012, vol.~30.

\end{thebibliography}

\end{document}